# A field-induced reentrant insulator state of a gap-closed topological insulator ($Bi_{1-x}Sb_x$) in quantum-limit states

Y. Kinoshita[1*], T. Fujita[1], R. Kurihara[1], A. Miyake[1], Y. Izaki[2], Y. Fuseya[3], M. Tokunaga[1]

[1]*The Institute for Solid State Physics, The University of Tokyo, Chiba 277-8581, Japan*

[2]*Department of Physics, Tokyo Institute of Technology, Tokyo 152-8551, Japan*

[3]*Department of Engineering Science, University of Electro-Communications, Tokyo 182-8585, Japan*

* kinoshita@issp.u-tokyo.ac.jp

In the extreme quantum limit states under high magnetic fields, enhanced electronic correlation effects can stabilize anomalous quantum states. Using band-tuning with a magnetic field, we realized a spin-polarized quantum limit state in the field-induced semimetallic phase of a topological insulator $Bi_{1-x}Sb_x$. Further increase in the field injects more electrons and holes to this state and results in an unexpected reentrant insulator state in this topological semimetallic state. A single-particle picture cannot explain this reentrant insulator state, reminiscent of phase transitions due to many-body effects. Estimates of the binding energy and spacing of electron-hole pairs and the thermal de Broglie wavelength indicate that $Bi_{1-x}Sb_x$ may host the excitonic insulator phase in this extreme environment.

**INTRODUCTION**

Magnetic fields are essential for studying condensed matter physics because they can continuously control the spins, orbital motion, and phases of electrons, which determine the nature of materials. However, the effect of the artificially applied magnetic field is weak compared to other energy scales that govern the electric states in materials, such as the Coulomb and electron–phonon interaction. Therefore, they are treated as perturbations to existing states. Conversely, in the extremely high magnetic fields that are thought to exist in neutron stars (> $10^8$ T) [1], magnetic fields dominate the other interactions and may realize an unknown world of materials.

Semimetals, in which electron and hole carriers coexist, are suitable for studying the properties of a material under extremely high magnetic fields [2]. This electron–hole system in semimetals can be assimilated to an electron–proton system. The attractive Coulomb interaction between the electrons and holes results in the binding energy of $R_y^*$. If we measure the effect of the magnetic flux density ($B$) on the electron by the ratio between the cyclotron energy ($\hbar\omega_c$) and the effective Rydberg energy $R_y^*$, it is given by [2, 3]

$$\gamma \equiv \frac{\hbar\omega_c}{2R_y^*} = \frac{16\pi^2\hbar^3\kappa^2\varepsilon_0^2}{\mu^{*2}e^3}B = 4.25 \times 10^{-6}\ [\text{T}^{-1}] \times \frac{m_0^2\kappa^2}{\mu^{*2}}B\ [\text{T}]. \quad (1)$$

Here, $\mu^*$ is the reduced mass defined as $(\mu^*)^{-1} = (m_e^*)^{-1} + (M_h^*)^{-1}$, where $m_e^*$ ($M_h^*$) denotes the effective mass of the electron (hole); $m_0$ is the mass of the free electron; $e$ and $\hbar$ are the elementary charge and the reduced Planck's constant, respectively; $\kappa$ and $\varepsilon_0$ are the relative dielectric constant and the dielectric constant of vacuum, respectively. In elemental bismuth with $\mu^* \sim 10^{-2}$ free electron mass ($m_0$) and $\kappa \sim 100$ [4], the effect of the magnetic field is $\sim 10^8$ times larger than that of the electron–proton system ($\kappa = 1$, $\mu^*$

= $m_0$). In the current high-field studies on BiSb alloys, we investigate the effect of continuously increasing the density of positive and negative charges from zero under the extremely high-field conditions.

Before discussing the properties under magnetic fields, we briefly review the electronic structures of pure and Sb-substituted bismuth at zero fields. Elemental bismuth is a semimetal with three electron pockets at the *L* points and a hole pocket at the *T* point in the first Brillouin zone. At the *L* point, the bands of different parities face each other across a small gap; the topology changes depending on whether there is a band inversion at this point. Currently, the topology of pure Bi is being extensively studied [5-10]. According to historical interpretations, it is topologically trivial [11]. Partial substitution of Bi with Sb results in band inversion at an Sb concentration of approximately 4%, making this material topologically non-trivial. Above 4%, a direct gap opens between the valence band and the conduction band of the electrons at the *L* point. With further substitution, the overlap between the valence band of the holes at the *T* point and the conduction band of the electrons at the *L* point disappears at approximately 7%, as shown in Fig. 1(a). Above this semimetal–semiconductor transition critical concentration, the system becomes a topological insulator [11].

The application of external magnetic fields can control similar band shifts in a valley-selective manner. When Landau quantization occurs in a magnetic field, the energy shifts of the Landau sub-bands are determined by the competition between the cyclotron and Zeeman energies. According to the theory of parameter optimization in 2017 [12], applying a field perpendicular to the trigonal axis of the semimetallic Bi crystal first causes band inversion at the *L* point. Further increase of the field leads to valley polarization due to anisotropy in the cyclotron masses and g-factors [12,13]. Finally, all carriers are eventually depopulated at the semimetal–semiconductor transition field.

Conversely, magnetic fields along the trigonal axis increase the band overlap because the Zeeman energy dominates the cyclotron energy at the highest Landau level for holes at the *T* point. Therefore, in a topological insulator BiSb alloy, applying a magnetic field along the trigonal axis causes a semiconductor–semimetal transition, as illustrated in Fig. 1(b) [14-16]. In this case, overlapping valence and conduction bands are both spin-split lowest/highest Landau sub-bands: this field-induced semimetallic state is in a spin-polarized quantum limit state. At low temperatures, we can continuously increase the density of electrons and holes or the Fermi energy, $E_F$, measured from the band top/bottom from zero by increasing the magnetic field from the transition field. Our present high-field study revealed the emergence of an insulating state in the high-field limit of $E_F \ll \hbar\omega_c$ at the equilibrium state.

**METHODS**

A single crystal of 9.9%-doped Bi–Sb alloy prepared using the zone melting method was used as the sample. Sb concentration was analyzed using an electron probe microanalyzer (JEOL JXA-8100). Several sample pieces were measured in the experiments, and the details are provided in Supplemental Material S1. For example, sample pieces named #1b and c were cut using a spark cutter from single piece named #1a, which was the identical piece measured in [16]. The sample sizes of #1a, #1b, and #1c were 1.5 × 2.0 × 3.4 mm³, 1.5 × 0.7 × 3.0 mm³, and 1.5 × 0.7 × 0.3 mm³, respectively.

The magnetotransport properties were measured using a four- or five-probe method and the numerical AC lock-in technique [17] or the DC technique. A non-destructive pulse magnet installed at the ISSP was used to generate pulsed high magnetic fields up to 60 T with duration of 36 ms.

For the ultrasonic measurement, we adopted an ultrasonic pulse-echo method with a numerical vector-type phase detection technique to measure the longitudinal ultrasonic velocity, $v_{33}$ [18]. The elastic constant was calculated as $\Delta C_{33}/C_{33} = 2\Delta v_{33}/v_{33}+(\Delta v_{33}/v_{33})^2$. To generate longitudinal ultrasonic waves, LiNbO$_3$ plates were used as piezoelectric transducers.

The band structures of Bi$_{1-x}$Sb$_x$ ($x = 0.096$ at $B = 0$ T) were calculated based on the Liu–Allen tight-binding model [19] with the virtual crystal approximation, where the tight-binding parameters were evaluated by a linear interpolation between pure Bi and Sb [20]. The Sb concentration, $x$, was scaled up by a factor of 2 to coincide with the experimental results (the band inversion at the $L$-point occurs at $x=0.04$) [20,21]. We calculated the Landau level spectrum using the nonperturbative matrix mechanics method ($\pi$-matrix) [22]. In the calculation of Fig. 6, we neglected the contributions of bare free electrons, which gives trivial contributions, to make the analysis as clear as possible.

**RESULTS AND DISCUSSION**

Figure 1(c) shows the temperature ($T$) dependence of the resistivity along the trigonal axis ($\rho_{zz}$) of the Bi$_{1-x}$Sb$_x$ alloy ($x = 0.099$) at zero magnetic field. The measured sample was #1b, and it was obtained from #1a (Methods and Supplemental Material). $\rho_{zz}$ increases upon cooling to approximately 20 K and it almost saturates below this temperature. This behavior is qualitatively consistent with that reported for semiconducting Bi$_{1-x}$Sb$_x$ alloys [21]. We evaluated the activation-type energy gap $\Delta \sim 15$ meV using the $\rho_{zz}$-$T$ curve above 20 K with $\rho_{zz}(T) = \rho_{zz}(0)\exp(-\Delta/2k_\text{B}T)$, as shown in the inset. According to [15], the estimated gap in the $\rho_{zz}$-$T$ curve coincides with the direct gap $E_L$ at $L$ point determined from the optical measurements. Therefore, $E_L$ was ~ 15 meV, as shown in Fig. 1(b).

Figure 2(a) presents the results of longitudinal magnetoresistance, $\rho_{zz}$ ($B \parallel I \parallel$ trigonal axis), at $T = 1.4$ K measured in #1a. There is a peak of approximately 70 mΩ•cm at approximately 2 T, and it decreases rapidly with increasing magnetic field. This behavior is consistent with the results measured on an identical sample piece in [16], indicating that the sample is not degraded [23]. The decrease in $\rho_{zz}$ has been regarded as a semiconductor-semimetal transition [16], whereas a recent study argued that the negative longitudinal magnetoresistance stems from the chiral anomaly [24]. Regardless of its origin, Fenton *et al*. confirmed that the negative magnetoresistance effect in this field region is less prominent with sample degradation [15]. In our sample, ratio $R = \rho_{zz}(4.2$ K, 10 T$) / \rho_{zz}(300, 0$ T$) = 27$, which is close to the lowest value, that is, the highest quality samples studied in [15]. The results for the longitudinal magnetoresistance in the samples with $R = 43$ and 95 are presented in Fig. S1 of the Supplemental Material which clearly shows that the magnitude of the negative magnetoresistance effect decreases with increasing $R$.

In this study, we observed an additional peak structure of $\rho_{zz}$ at a higher field. For convenience, we refer to the peak structures centered around 2 T and 20 T as the first and second peaks, respectively. Although we can observe another peak-like structure at a positive field of approximately 15 T (red line), it is hardly visible after symmetrizing, $[R(+B) + R(-B)]/2$, for the data in positive and negative magnetic fields (green line). Therefore, we interpret this structure to originate from the Hall resistivity, $\rho_{yx}$, component owing to misalignment of the terminals.

The results of the transverse magnetoresistance at 1.4 K under applied fields along the trigonal axis are presented in Figs. 2(b) and (c). Because we used the same sample piece, #1a, as that for $\rho_{zz}$ measurements, there may be quantitative uncertainty in the absolute resistivity value owing to the unsuitable shape of the sample for transport measurements, as illustrated in the inset (Supplemental Material S4). A more reliable data set is discussed later. $\rho_{xx}$ exhibits a peak below 10 T and it decreases monotonically with increasing magnetic field. The simultaneously measured Hall resistivity, $\rho_{yx}$, exhibits a peak structure at approximately 12 T and then it approaches zero. We did not observe quantum oscillations in either $\rho_{xx}$ or $\rho_{yx}$, contrary to the results reported for similar BiSb alloys [25], but similar to that in Ref.24.

Figure 3(a) shows the longitudinal magnetoresistance, $\rho_{zz}$, at various values of $T$. The first and second peaks are suppressed with increasing temperature. The difference between them is apparent when the peak values are plotted as a function of the temperature [inset of Fig. 3(a)]. Both peaks exhibit semiconducting negative temperature dependence. Contrary to the monotonic increase in $\rho_{zz}$ at the first peak upon cooling, the second peak increased steeply below approximately 5 K. This behavior indicates a phase transition induced below the critical temperature. Figure 3(b) shows $1/T$ dependence of $\log\rho_{zz}$ at 20 T in the second peak region. The black dashed line shows the fitting result using the activation-type function, $\exp(\Delta/k_B T)$ and $\Delta$ is evaluated approximately 0.8 meV. Above approximately $1/T = 0.5$ K$^{-1}$ ($T = 2$ K), the second peak saturates and deviates from the black dashed line. The overall behavior of $\rho_{zz}$ in the $T$-$B$ plane can be observed in the contour plot shown in Fig. 3(c). The orange and red dots represent the points where $\rho_{zz}$ crosses 0.5 mΩ·cm at the first and second peaks, respectively. The second peak shows a dome-like high-resistance region after suppressing the insulating state below 10 T using a magnetic field.

Although transport measurements can sensitively detect any changes in electronic states, it is difficult to discuss the background physics from these data alone. Thus, we studied ultrasonic propagation, which reflects the thermodynamic properties of electronic systems in magnetic fields applied along the trigonal direction. The red line in Fig. 4(a) shows the field dependence of the relative change in longitudinal elastic constant $C_{33}$ propagating and polarizing along the trigonal axis. $C_{33}$ was almost flat up to approximately 10 T. In contrast, $C_{33}$ showed a kink structure at approximately 11 T and then it decreased monotonically. Assuming a rigid two-band model with an identical constant density of states, the elastic constant, $C$, can be expressed according to the following equation [26]:

$$C = C^0 - 2d^2 N_0 \left[1 - e^{-\frac{E_F}{k_B T}}\right], (2)$$

where, $C^0$ is the background elastic constant, $d$ is the deformation potential due to the crystal strain, $E_F$ is the Fermi energy measured from the top (bottom) of the valence (conduction) band, and $N_0$ is the density of states at $E_F$. The field-independent behavior of $C_{33}$ is well described in Eq. (2) because the density of state $N_0$ is zero in the insulating state. In contrast, the experimentally observed kink-like decrease in $C_{33}$ can be ascribed to the steep increase in $N_0$, which is consistent with the emergence of a semiconductor–semimetal transition at $B_c$ = 11 T. In addition to the kink anomaly at $B_c$, a weak but significant change in the $C_{33}$ ($B$) slope was observed at approximately 27 T, as indicated by the yellow arrow.

Additionally, we examined how the anomalies observed in the ultrasonic measurements correspond to the changes in the magnetotransport properties. Figure 4(b) shows the magnetic field dependence of $\rho_{xx}$ and $\rho_{yx}$ of sample (#1c) cut from sample #1a into a shape suitable for transport measurements. The black line denotes $\rho_{zz}$, which is the same as the green line in Fig. 2(a). $\rho_{xx}$ (red line) and $\rho_{yx}$ (blue line) show a peak and kink at approximately $B_c = 11$ T. From the data of $\rho_{xx}$, $\rho_{yx}$, and $\rho_{zz}$, we evaluated the anisotropy ratio of conductivity, $\sigma_{zz}/\sigma_{xx}$, as

$$\frac{\sigma_{zz}}{\sigma_{xx}} = \frac{\rho_{xx}^2 + \rho_{yx}^2}{\rho_{xx}\rho_{zz}}. \quad (3)$$

As shown by the green line in Fig. 4(a), the ratio exceeds 100 in the semimetallic phase above 11 T, indicating that the system is a quasi-one-dimensional conductor. This quasi-one-dimensionality is suppressed at approximately 20 T and it is revived at higher magnetic fields. Although $\sigma_{zz}/\sigma_{xx}$ exhibits two peaks at 14 and 16 T, this structure originates from the slight Hall resistivity, $\rho_{yx}$, component on $\rho_{zz}$, which cannot be eliminated completely by symmetrizing.

Field-induced semiconductor–semimetal transition has been anticipated because the large Zeeman effect dominates over the cyclotron energy in the hole band of this material. Because the ratio of these two energies shows characteristic anisotropy, we can determine whether $B_c$ represents a semiconductor–semimetal transition through its field-angle dependence. Therefore, we investigated transverse magnetoresistance as a function of the magnetic field direction.

Figures 5(a) and (b) show the field-angle dependence of $\rho_{xx}(B)$ and $\rho_{yx}(B)$ measured simultaneously at 1.4 K. Here, electric currents were applied along the binary direction, and magnetic fields were tilted by an angle $\theta$ measured from the trigonal to the bisectrix axis. The peaks in $\rho_{xx}$ and kinks in $\rho_{yx}$ (shown by red and blue arrows, respectively) systematically move to higher magnetic fields with increasing $\theta$. Red and blue circles in Fig. 5(c) represent these characteristic magnetic fields as a function of $\theta$.

Additionally, we compared the result with that expected for the semiconductor–semimetal transition. Because the electrons at the $L$ point can be regarded as Dirac electrons, we assume that the lowest Landau sub-band is less sensitive to the applied magnetic field because of the cancelation between the Zeeman and cyclotron energies [27]. Therefore, the indirect gap, $E_g$, collapses because of the increase in energy of the highest hole Landau sub-band: transition occurs when $E_g = \frac{1}{2}(g^*\mu_B B - \hbar\omega_c)$. Therefore, the transition field is given as

$$B_c(\theta) = \frac{2M^*(\theta)E_g}{\hbar e[M_{ZC}(\theta)-1]}, \quad (4)$$

where, $M_{ZC}(\theta)$ and $M^*(\theta)$ denote the ratio between the Zeeman and cyclotron energy and cyclotron mass of holes at field angle $\theta$, respectively. Using eq. (40) and (41) in [27], we can estimate $M^*(\theta)$ as

$$M^*(\theta) = \sqrt{\frac{\det \hat{M}}{M_h}} = \left[\left(\frac{\cos\theta}{M^*_{XY}}\right)^2 + \left(\frac{\sin\theta}{M^*_{ZX}}\right)^2\right]^{-1/2}, \quad (5)$$

$$M_h = \mathbf{h} \cdot \hat{M} \cdot \mathbf{h}, \quad (6)$$

where $\hat{M}$ is the mass tensor and $\mathbf{h} = (0, \sin\theta, \cos\theta)$ is a unit vector along the magnetic field. $M^*_{XY}$ and $M^*_{ZX}$ are the cyclotron masses for the magnetic fields applied in the trigonal and bisectrix directions, respectively. We used the values $M^*_{XY} = 0.0678\ m_0$ and $M^*_{ZX} = 0.221\ m_0$ for pure bismuth in the BiSb alloy [28]. Because the $M_{ZC}(\theta)$ of this sample is unknown, we used the experimental values reported for pure bismuth [29]. Thus, the calculated $B_c(\theta)$ is plotted as a dotted line in Fig. 5(c). By tuning a single adjustable parameter, $E_g = 10.5$ meV, we can reasonably reproduce the experimental data using eq. 5. Therefore, we ascribe $B_c$ as a critical field for the semiconductor–semimetal transition caused by crossing the spin-split electron and hole Landau subbands with the indices $N = 0$.

Contrary to the systematic angular variation of $\rho_{xx}$, the second peak in $\rho_{zz}$ was hardly discernible by a slight change in $\theta$, as shown in the inset of Fig. 5(d). Figure 5(e) shows the peak value of $\rho_{zz}$ as a function of $\theta$, which shows peaks only if the field is applied close to the trigonal direction.

Subsequently, we consider the origin of the second peak in $\rho_{zz}$. The presence of twinning in Bi crystals leads to the appearance of extra peaks in quantum oscillations [29]. Our experiments showed a systematic field-angular dependence of $B_c$, suggesting the single-domain nature of the crystal. Even in the presence of twins, a change in the resistivity of the secondary domains is unlikely to cause an increase in resistivity in the semimetallic state above $B_c$.

Another possibility is the decrease in carriers owing to magnetic freeze-out (e.g. $n$-type InSb [30]), that is, the carriers in the magnetic fields are localized within the magnetic length, $l_B = \sqrt{\hbar/eB}$, and trapped by the impurity ions. Therefore, in this phenomenon, the resistance increases monotonically and exhibits isotropic behavior. On the other hand, only $\rho_{zz}$ showed a reentrant insulator phase and a dome-like phase diagram [Fig. 3(c)]. Therefore, magnetic freeze-out cannot be the origin of the second peak.

Fenton *et al.* observed a hump structure on the high-field side of the peak at $\rho_{zz}$ below 10 T for a semiconducting BiSb alloy [15]. They ascribed this to the lift of the three-fold degeneracy of the electron valleys due to misalignment of the field. As shown in Figs. 5(d) and 5(e), the second peak becomes prominent when the applied magnetic field is parallel to the trigonal direction. By contrast, the hump structure in [15] emerged when the applied fields were slightly tilted from the trigonal axis.

Hiruma *et al.* observed a similar hump structure in $\rho_{zz}$ and regarded it as crossing other Landau sub-bands [16]. To clarify the Landau level spectrum in this material, we calculated it using the $\pi$-matrix method with virtual crystal approximation (VCA) based on the Liu–Allen's tight binding model [19,22]. The Landau level spectrum for $B \parallel$ trigonal axis for the Sb substitution that gives $E_L = 15.2$ meV is shown in Fig. 6, which js consistent with our evaluated activation energy for $x = 0.099$. The cyclotron masses of the electrons and holes for the $B \parallel$ trigonal axis are $m_{ce}/m_0 = 0.0176$ and $M_{ch}/m_0 = 0.0974$, respectively. The highest Landau level of holes (0h-) at $T$ point increases with increasing magnetic field, whereas that of the lowest Landau level of electrons (0e-) at $L$ point decreases. They cross each other at $B_c = 11$ T. (The energy of the hole band is decreased by 39 meV from the value of simple VCA.) The highest Landau level of the valence band at the L-point increases with the magnetic field, resulting in the lowest Landau level inversion at approximately $B_{c2} = 27$ T, which was obtained without any adjusted parameters. This level crossing may appear as a bending of the elastic constant in Fig. 4(a), but we did not observe any anomalies in the magnetotransport properties. From the calculated results, we estimated the Fermi level for several $B$ (Supplemental Material S7) as shown in Fig. 6 as green dots. It exists near the highest Landau level of holes (0h-), which is far from the level crossing point. In this situation, it is not surprising that level crossing causes no significant changes in the physical properties. In practice, band inversion at the $L$ points by Sb substitution is known not to show any anomalies in magnetic susceptibility [31, 32] which is theoretically explained in Ref. 33.

The theoretical calculation confirmed that the field-induced semimetallic state above 11 T is in the spin-polarized quantum limit state. In this state, all electrons and holes occupy only the spin-split Landau levels with $N = 0$. Recently, anomalous magnetotransport properties in the quantum limit state have attracted significant attention for several topological materials [34-38]. In the present material, coexisting electrons and holes in the compensated semimetal may cause a phase transition that is peculiar to such systems.

In the following, we will discuss the possibility of realizing of an excitonic phase [39-41], which has long been expected to emerge in the quantum limit state of the electron–hole systems [42].

According to conventional arguments, electron–hole pairs, i.e. excitons, are spontaneously formed when the bandgap in a semiconductor or the band overlap in a semimetal becomes smaller than the binding energy of the exciton. The excitonic phase is the quantum condensation state of excitons at low temperatures. The excitonic phase is expected to be Barden–Cooper–Schrieffer (BCS)-type in the semimetal region and Bose–Einstein condensation (BEC)-type in the semiconductor region [43]. The transition temperature can be the highest in the crossover region. Although a dome-like excitonic phase boundary has been reported in semimetallic graphite at high fields [44], there is no clear experimental evidence of BEC/BCS crossover near the semimetal–semiconductor transition. On the other hand, BEC/BCS crossover has been extensively studied in experiments on cold atoms [43]. In this case, crossover occurs when the size of the boson spatial extension and the interparticle distance become comparable.

Here, we will discuss our experimental results from this perspective. The field-induced insulating state was the most stable in the semimetallic region, as shown in Fig. 3(c). First, we estimate the exciton binding energy ($E_B$) in this field region. According to [45] and [46], $E_B$ monotonically increases as $\gamma$ in eq. 1 increases in proportion to the magnetic field. For the present material at $B = 20$ T, $\gamma$ and $E_B$ are estimated to be ~4000 and ~0.3 meV (Supplemental Material S5), respectively. For hydrogen atoms, that is, an electron–proton system, $\gamma = 4000$ can be realized at $B = 1 \times 10^9$ T. At such a high-field limit, the formation of excitons by the enhanced $E_B$ is possible in the $B$-$T$ region where the second peak is observed. In addition, the effective Bohr radii of excitons decrease under high magnetic fields in both the longitudinal and transverse directions [45]. For $\gamma = 4000$, the longitudinal radii decrease to ~0.15 $a_B^*$ (Supplemental Material S5), where $a_B^*$ is the effective Bohr radius at zero field given as

$$a_B^* = \frac{4\pi\kappa\varepsilon_0\hbar^2}{\mu^* e^2}. \quad (7)$$

We calculated $\mu^*$ as $\mu^* \sim 0.0117\ m_0$ using the band mass of $Bi_{1-x}Sb_x$ ($x = 0.096$ at $B = 0$ T) as $m_e^*/m_0 = 0.0125$ and $M_h^*/m_0 = 0.184$, each of which was calculated using the geometric mean of the diagonalized band mass tensor (Supplemental Material S6). Using the calculated $\mu^*$ and the typical parameters of elemental bismuth as $\kappa \sim 100$, $a_B^*$ was approximately 450 and 70 nm at $B = 0$ T and 20 T, respectively.

Conversely, the spacing of the carriers confined in a Landau tube in the quantum limit state can be given by the Femi wavelength $\lambda_F$ as

$$\lambda_F = \frac{2\pi}{k_F^\hbar}. \quad (8)$$

The Fermi wavenumbers (Fermi energy) for electrons $k_F^e$ ($E_F^e$) and holes $k_F^h$ ($E_F^h$) vary with $B$. The values at 20 T were estimated from the calculated $B$-dependence of the band overlap $E_0$ (Fig. 6) and the dispersion relation along $B$ (Fig. S5) under the conditions $3k_F^e = k_F^h$ and $E_0 = E_F^e + E_F^h$. The former constraint originates from charge neutrality and three-fold electron valley degeneracy in the quantum-limit state. Thus, the estimated $\lambda_F$ was approximately 54 nm at 20 T. Therefore, the BEC/BCS crossover field can be estimated to coincide with the apex of the dome structure, as shown in Fig. 3(c).

Additionally, we estimated the temperature of BEC ($T_{BEC}$). $T_{BEC}$ can be defined as the point at which the thermal de Broglie wavelength $\lambda_T = h/\sqrt{2\pi m^* k_B T}$ becomes comparable to the interparticle distance, $\lambda_F$, where $h$ is the Planck's constant and $k_B$ is the Boltzmann constant. The mass of the exciton is $m^* = 0.661\ m_0$, which is the sum of the calculated band masses along the trigonal axis of the electron and hole as $m_e^* = 0.00637 m_0$ and $M_h^* = 0.655\ m_0$, respectively. Using $m^*$, $\lambda_T$ becomes $\lambda_F$ (20 T) = 53 nm at $T = 3$ K, which is near the temperature where the 2nd peak emerges, 5 K.

All these estimations are consistent with the emergence of the excitonic phase, which has long been investigated in BiSb alloys. However, since we do not have direct evidence that excludes the other scenario at the present stage, the excitonic phase is not the only possible origin of the reentrant insulator state. We observed gap-like behavior in the longitudinal magnetoresistance, while no significant anomaly in the transverse configuration. Although we do not have a clear explanation of the origin of this anisotropic response, similar behaviors are commonly observed in quantum limit states of graphite [34], TaAs [36], and $CaIrO_3$ [38]. According to Halperin, gap-opening in the lowest Landau subband in the quantum limit state, e.g., by a density wave formation, can realize quantum Hall states in three-dimensional materials [47]. Tang *et al*. have claimed the realization of such a quantum Hall state in $ZrTe_5$ [37]. Fauqué *et al*. discussed the contribution of edge conduction to the quantum Hall state in the insulating phase of graphite in magnetic fields above 53 T. Surface-state conduction may explain the saturation of the resistance in the second peak at low temperature [Fig. 3(b)] and the decrease in anisotropy around 20 T [Fig. 4(a)] in the present results. In any case, the origin of gap-opening in the bulk state is an essential issue to be clarified. The thermodynamic properties of this insulating state will be elucidated through detailed studies using more sensitive probes in the future. In addition, since many of topological semimetals are in the vicinity of the zero-gap state, the study of the gap-closed quantum limit will bring new developments in the physics of topological materials.

## CONCLUSION

In this study, we investigated the transport properties of 9.9%-doped Bi–Sb alloys in the quantum limit state using magnetoresistance and ultrasonic measurements in magnetic fields up to 60 T. The results of magnetotransport and elastic measurements indicate the emergence of a semiconductor–semimetal transition at a field of 11 T applied along the trigonal axis. A further increase in the magnetic field induced an additional insulating state in the spin-polarized quantum limit state at approximately 20 T. The calculated shifts in the Landau sub-bands cannot explain the emergence of the insulating state in a single particle picture. Therefore, the many-body effect is expected to play a crucial role in the insulating state.

## ACKNOWLEDGEMENTS

We thank T. Yamanaka and H. Yaguchi for their assistance with the electron probe microanalyser. We also thank K. Hiruma and N. Miura for the providing the $Bi_{1-x}Sb_x$ samples. This work was supported by Grant-in-Aids for Scientific Research nos. JP19H01850 and 21K04889 from MEXT.

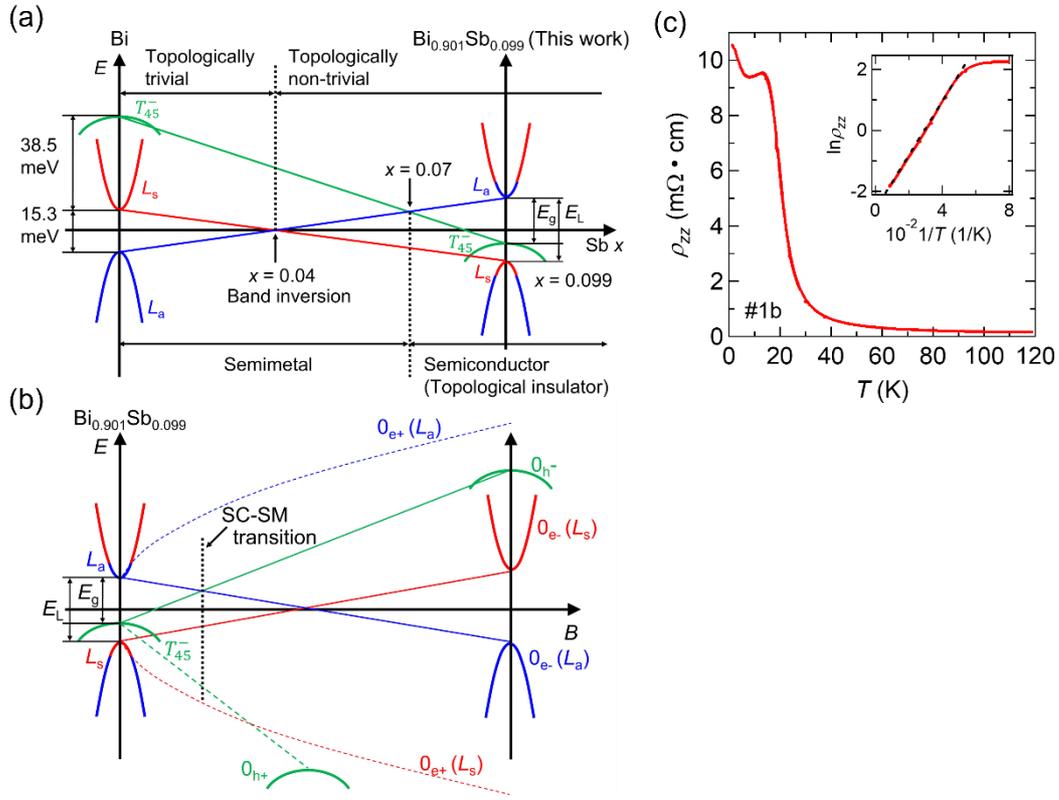

FIG. 1. (a) Schematic diagram of the energy bands of Bi with Sb concentration. (b) Schematic diagram of the energy bands with applying a magnetic field to BiSb alloys. (c) Temperature dependence of resistivity $\rho_{zz}$ at zero field measured on the sample piece #1b. The inset shows the Arrhenius plot of $\rho_{zz}$ as red line. The black dotted line is the fitting result by $\rho_{zz}(T) = \rho_{zz}(0)\exp(-\Delta/2k_\mathrm{B}T)$.

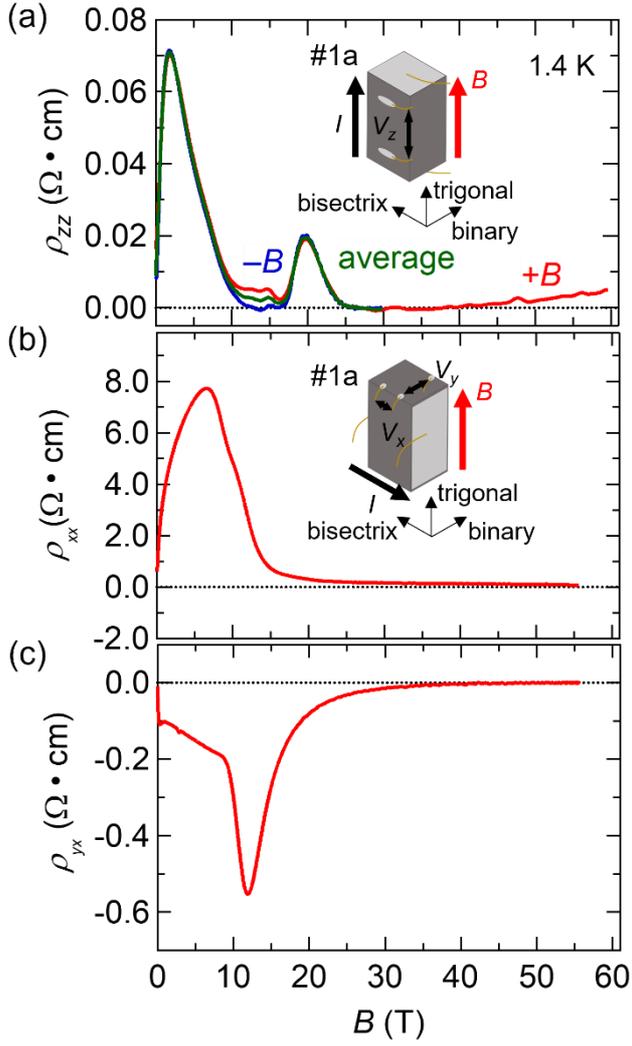

FIG. 2. Magnetoresistivity up to ~60 T at 1.4 K. The measured sample piece is #1a. (a) Longitudinal magnetoresistivity $\rho_{zz}$ in the configuration of $B \parallel I \parallel$ trigonal axis. The red(blue) line shows the result in the positive(negative) magnetic fields and the symmetrized data by both results is shown as the green line. (b) Transverse magnetoresistivity, $\rho_{xx}$, and Hall resistivity $\rho_{yx}$ in the configuration of $B \parallel$ trigonal axis, $I \parallel$ bisectrix axis.

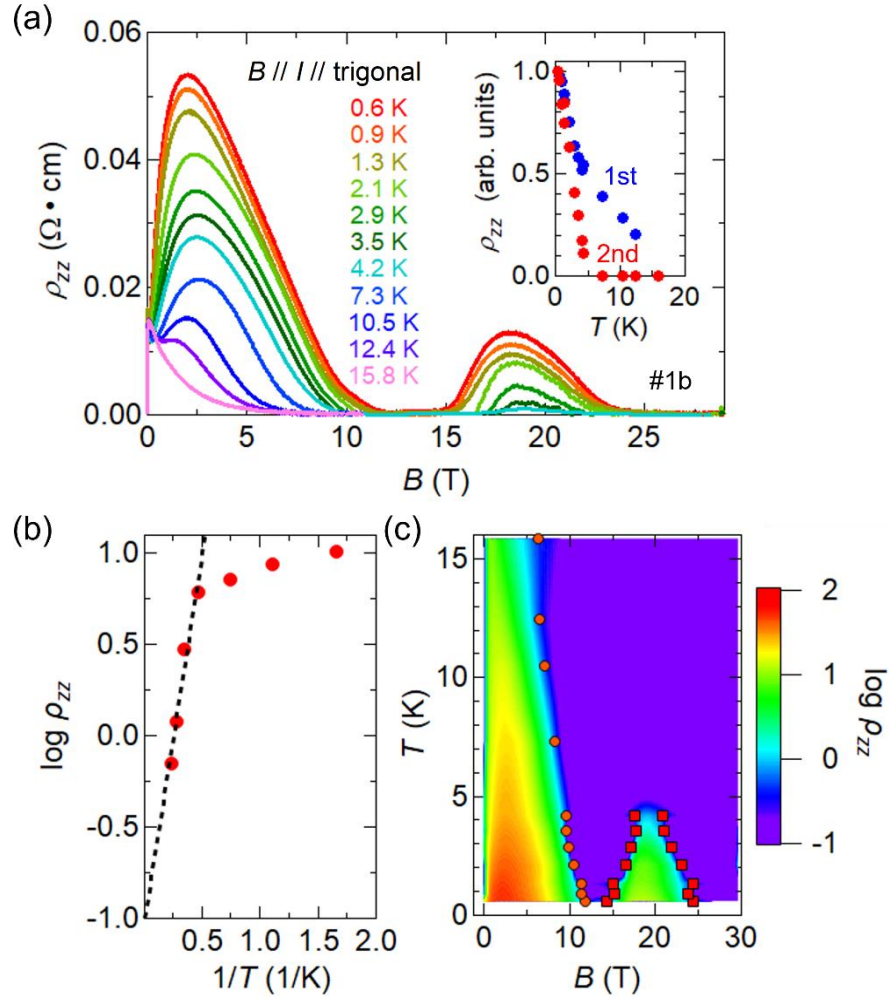

FIG. 3. (a) Longitudinal magnetoresistivity $\rho_{zz}$ at various temperatures in the configuration of $B \parallel I \parallel$ trigonal axis. The sample piece is #1b. The inset shows temperature dependence of the maximum $\rho_{zz}$ of each peak. The values of each peak are normalized by the maximum value at 0.5 K. (b) $1/T$ dependence of the $\log\rho_{zz}$ at 20 T. The dashed black line shows the fitting result by activation-type function, $\exp(\Delta/k_B T)$. (c) The contour plot of $\log\rho_{zz}$ as a function of temperature and magnetic field in the configuration of $B \parallel I \parallel$ trigonal axis. The orange (red) dots represent the points where $\rho_{zz}$ crosses 0.5 m$\Omega\cdot$cm at the 1st (2nd) peak.

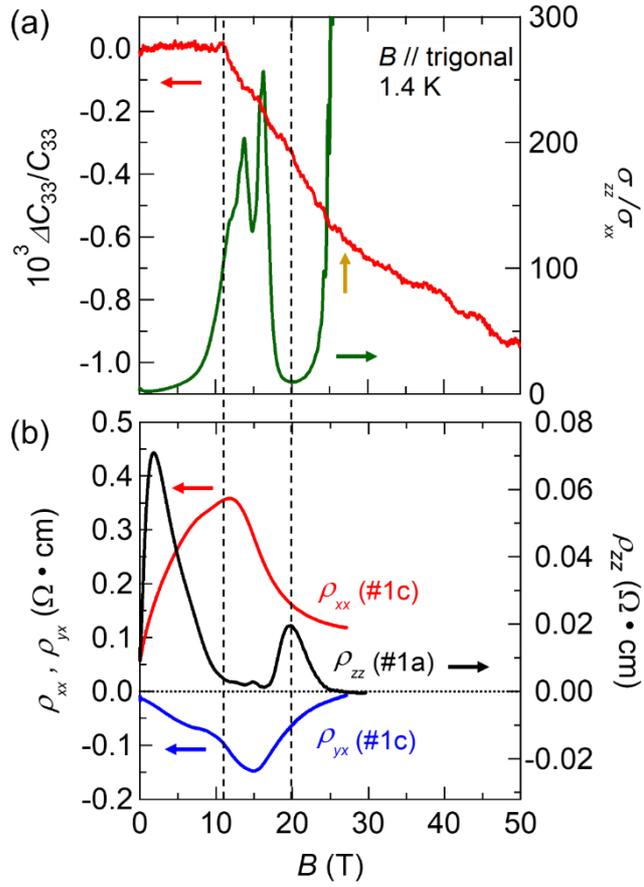

FIG. 4. (a) The relative change in the elastic constants $\Delta C_{33}/C_{33}$ (red line) and the ratio of conductivity $\sigma_{zz}/\sigma_{xx}$ (green line) versus the magnetic field parallel to the trigonal axis (b) Magnetoresistivity $\rho_{xx}$ (red line), $\rho_{yx}$ (blue line), and $\rho_{zz}$ (black line) at 1.4 K. The $\rho_{xx}$, $\rho_{yx}$ were measured on the #1c sample in the configuration of $B \parallel$ trigonal axis, $I \parallel$ binary axis. The $\rho_{zz}$ is the identical data with shown in Fig. 1(a) (green line).

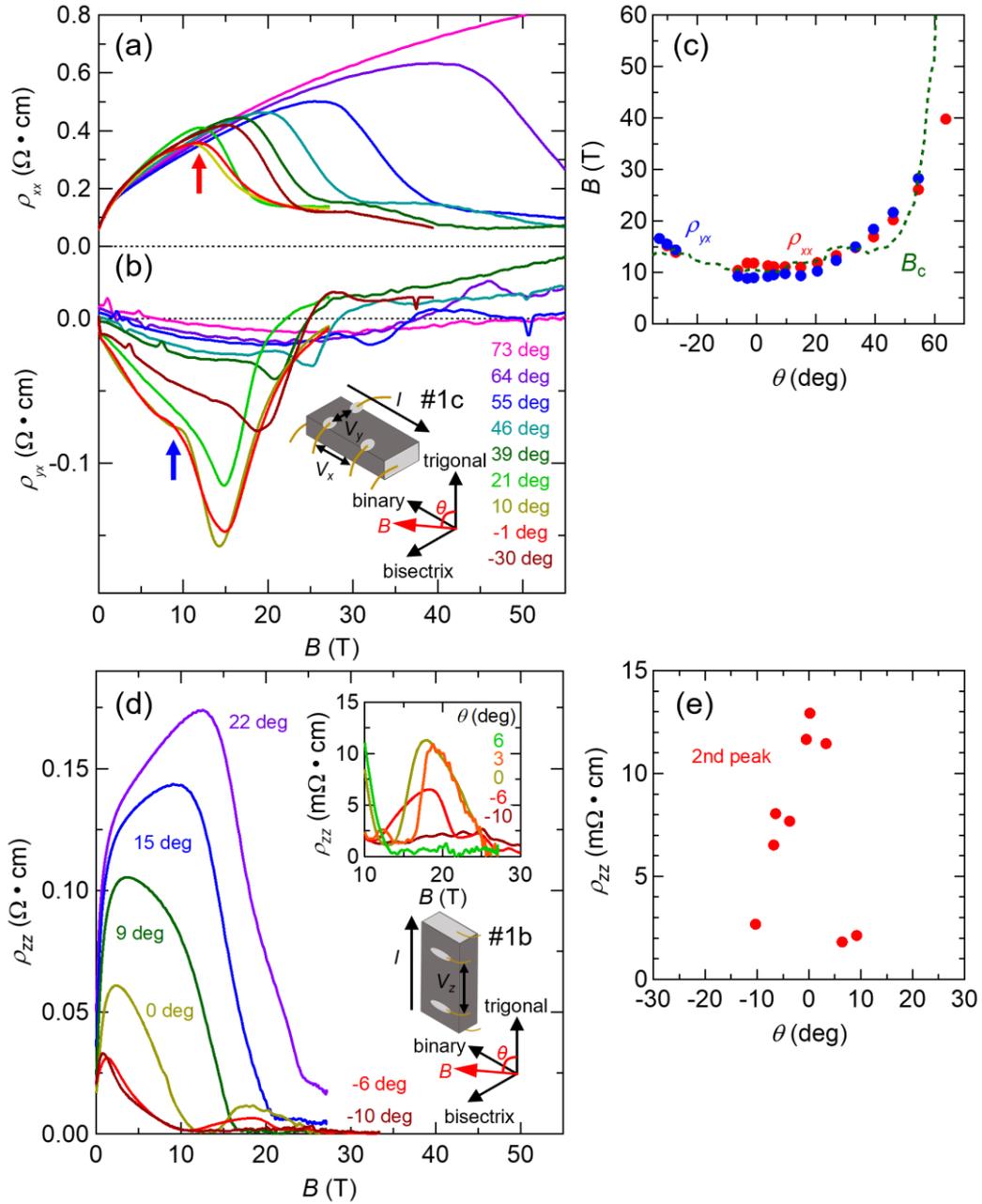

FIG. 5. (a) Magnetoresistivity $\rho_{xx}$, and (b) $\rho_{yx}$ for various $\theta$ with $I \parallel$ binary axis at 1.4 K. $\theta$ is the angle of magnetic fields from the trigonal axis. The measured sample piece #1c was rotated in the binary plane. (c) The $\theta$ dependence of the peak (kink) position for $\rho_{xx}$ ($\rho_{yx}$) shown by red (blue) circles and the calculated transition field $B_c(\theta)$ plotted by a green dotted line. (d) Magnetoresistivity $\rho_{zz}$ for various $\theta$ measured on the #1b piece with $I \parallel$ trigonal axis at 1.4 K. The sample was rotated in the binary plane. The top right inset

shows $\rho_{zz}$ around the 2nd peak for several $\theta$ near the trigonal axis. (e)Field-angle dependence of the maximum of the 2nd peak of $\rho_{zz}$.

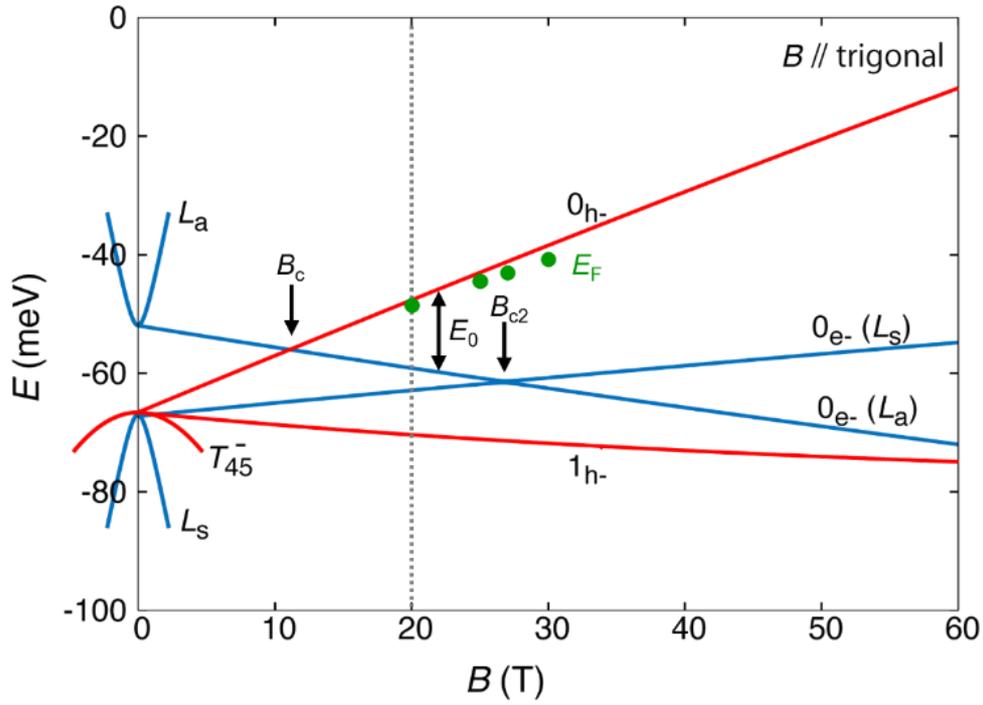

FIG. 6. Landau level spectrum for $B \parallel$ trigonal axis in $Bi_{1-x}Sb_x$ ($x \simeq 0.1$) calculated by the $\pi$-matrix method with the VCA based on Liu-Allen's tight binding model. The Sb concentration $x$ is determined so as to be $E_L \simeq 15$ meV. $0_{e-}$ indicates the lowest Landau level for electrons at the $L$-point. $0_{h-}$ and $1_{h-}$ indicate the highest and next-highest Landau levels for holes at the $T$-point.